\documentclass[11pt,a4paper]{ivoa}
\newif\iftth
\iftth
\newcommand{\ivoaDocversion}{1.0}
\newcommand{\ivoaDocdate}{2017-05-17}
\newcommand{\ivoaDocdatecode}{20170517}
\newcommand{\ivoaDoctype}{REC}
\newcommand{\ivoaDocname}{SODA}

\input tthntbib.sty

\begin{html}
<meta http-equiv="Content-type" content="text/html;charset=UTF-8"/>
\end{html}

\definecolor{ivoacolor}{rgb}{0.0,0.318,0.612}

\renewcommand{\author}[2][0]{\def\@tmp{#1}
  \if 0\@tmp
	{\begin{html}<li class="author">\end{html}#2\begin{html}</li>\end{html}}\else
	{\begin{html}<li class="author"><a href="#1">\end{html}#2\begin{html}</a></li>\end{html}}\fi}
\renewcommand{\previousversion}[2][0]{\def\@tmp{#1}
  \if 0\@tmp
	{\begin{html}<li class="previousversion">#2</li>\end{html}}\else
	{\begin{html}<li class="previousversion">
	  <a href="#1">#2</a></li>\end{html}}\fi}
\renewcommand{\ivoagroup}[1]
  {\begin{html}<dd id="ivoagroup">#1</dd>\end{html}}
\renewcommand{\editor}[2][0]{\def\@tmp{#1}
  \if 0\@tmp
        {\begin{html}<li class="editor">\end{html}#2\begin{html}</li>\end{html}}\else
        {\begin{html}<li class="editor"><a href="#1">\end{html}#2\begin{html}</a></li>\end{html}}\fi}

\newcommand{\includeMeta}{%
   \ivoaDocversion\ivoaDoctype\ivoaDocname\ivoaDocdate}

\def\SVN$#1: #2 ${%
	#2}

\newenvironment{abstract}{%
  \includeMeta
  \begin{html}
    </div> <!-- titlepage -->
    <div id="abstract"><h2>Abstract</h2>
  \end{html}
  }{%
    \ivoaDoctype
    \tableofcontents
  }

\newcommand{\lstloadlanguages}[1]{}
\newcommand{\lstset}[1]{}

\newenvironment{lstlisting}[1][plain]
  {\tthverbatim{lstlisting}}
  {}

\newcommand{\specialterm}[2]{%
  \begin{html}<span class="#1">\end{html}#2\begin{html}</span>\end{html}}
\newcommand{\xmlel}[1]{\specialterm{xmlel}{#1}}

\newcommand{\sptablerule}{}

\newcommand{\harvarditem}[4][0]{%
  
  \if 0#1 \item[#2 (#3)]
  \else \item[#1 (#3)]\fi}
\newcommand{\harvardurl}[1]{\url{#1}}

\def\AtBeginDocument#1{\relax}
\def\pgfsyspdfmark#1#2#3{\relax}

\newbox\voidb@x
\def\@m{\relax}

\begin{html}
  <div id="titlepage">
\end{html}

\fi

\newcommand{\xtype}[1]{\texttt{#1}}
\newcommand{\ucd}[1]{\texttt{#1}}

\usepackage{listings}
\usepackage[utf8]{inputenc}

\title{IVOA Server-side Operations for Data Access}

\ivoagroup{Data Access Layer Working Group}

\author{Fran\c cois Bonnarel}
\author{Markus Demleitner}
\author{Patrick Dowler}
\author{Douglas Tody}
\author{James Dempsey}

\editor{Fran\c cois Bonnarel, Patrick Dowler}

\previousversion{PR-SODA-1.0-20160429}
\previousversion{WD-SODA-1.0-20151221}
\previousversion{WD-AccessData-1.0-20151021}
\previousversion{WD-AccessData-1.0-20140730}
\previousversion{WD-AccessData-1.0-20140312}

\SVN$Rev: 4112 $
\SVN$Date: 2017-06-02 18:50:37 +0200 (Fri, 02 Jun 2017) $
\SVN$URL: http://volute.g-vo.org/svn/trunk/projects/dal/SODA/SODA.tex $

\begin{document}

\begin{abstract}
This document describes the Server-side Operations for Data Access
(SODA) web service capability.  SODA is a low-level data access
capability or server side data processing that can act upon the data
files, performing various kinds of operations: filtering/subsection,
transformations, pixel operations, and applying functions to the data.  
\end{abstract}

\section*{Acknowledgments}
The authors would like to thank all the participants in DAL-WG discussions for
their ideas, critical
reviews, and contributions to this document.

Fran\c cois Bonnarel  acknowledges support  from  the  Astronomy  ESFRI  and Research Infrastructure Cluster – ASTERICS project,  funded by the European Commission under the Horizon 2020 Programme (GA 653477).

\section*{Conformance-related definitions}

The words ``MUST'', ``SHALL'', ``SHOULD'', ``MAY'', ``RECOMMENDED'', and
``OPTIONAL'' (in upper or lower case) used in this document are to be
interpreted as described in IETF standard RFC2119 \citep{std:RFC2119}.

The \emph{Virtual Observatory (VO)} is a
general term for a collection of federated resources that can be used
to conduct astronomical research, education, and outreach.
The \href{http://www.ivoa.net}{International
Virtual Observatory Alliance (IVOA)} is a global
collaboration of separately funded projects to develop standards and
infrastructure that enable VO applications.

\section{Introduction}
The SODA web service interface defines a RESTful web service for
performing server-side operations and processing on data before
transfer.

\subsection{The Role in the IVOA Architecture}

\begin{figure}
\centering

\includegraphics[width=0.9\textwidth]{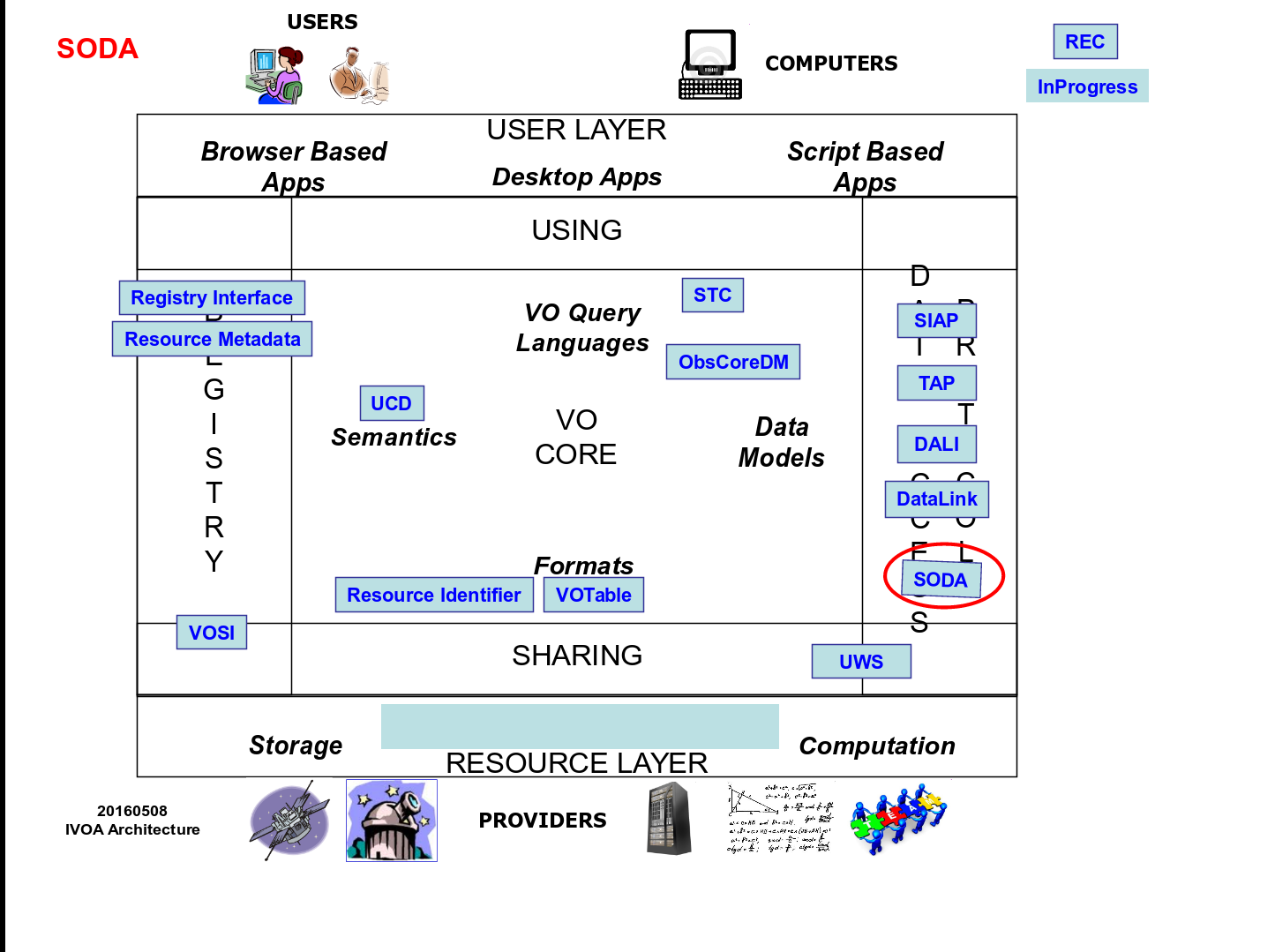}
\caption{SODA in the global VO architecture}
\label{fig:architecture}
\end{figure}

Figure~\ref{fig:architecture} shows how SODA fits into the IVOA architecture.
SODA services conform to the Data Access Layer Interface (DALI,
\citet{std:DALI}) specification, 
including the Virtual Observatory Support Interfaces (VOSI,
\citet{std:VOSI}) resources.

Within the IVOA architecture, SODA services could be found and used in
several ways. First, a SODA service could be found in the IVOA Registry
and used directly. A description of a SODA service may be found along
with specific dataset metadata at either the data discovery phase using
Simple Image Access (SIA, \citet{std:SIAv2}) or Table Access Protocol
(TAP, \citet{std:TAP}) and the ObsCore data model \citep{std:OBSCORE} or
via a DataLink \citep{std:DataLink} service. The service descriptors and
three-factor semantics rely on UCDs \citep{std:UCD} and the VO Unit
specification \citep{std:VOUNIT}. Since the discovery of SODA services
makes use of DataLink service descriptor(s) to provide parameter
metadata, the VOSI-capabilities specified in
Section~\ref{sec:capabilities} do not make use of a registry extension.

\subsubsection{SODA Service in the Registry}

Resources in the IVOA Registry may include SODA capabilities. In order to
use such services, clients require prior knowledge of suitable
identifiers that are usable with a registered SODA service. This
scenario is described in more detail below in
Section~\ref{sec:reg-soda}.

\subsubsection{SODA Service from Data Discovery}

In the simplest case, the identifiers found via data discovery can be
used directly with an associated SODA service. The query response (from
SIA or TAP) would include one or more service descriptor(s)
that describe the available SODA capabilities. This scenario is described
in detail in Section~\ref{sec:disc-soda}.

\subsubsection{SODA Service from DataLink}

In the general case, data discovery responses may direct clients to an
associated DataLink service where access details can be obtained. The
DataLink output will in turn provide service descriptor(s) of the
associated SODA service(s). Service providers may choose this approach
for several reasons; for example, one entry from data discovery may lead
to multiple files or resources, or access via services such as SODA may
be considered the primary access mode and direct download is not
available or discouraged. This scenario is described in detail in
Section~\ref{sec:disc-links-soda}.

\subsection{Motivating Use Cases}
Below are some of the more common use cases that have motivated the
development of the SODA specification. While this is not complete, it
helps to understand the problem area covered by this specification.

\begin{itemize}
\item{Retrieve Subsection of a Datacube}
\label{sec:use-cube}

Cutout a subsection using coordinate axis values. The input to the
cutout operation will include one or more of the following:

\begin{itemize}
\item a region on the sky
\item an energy value or range
\item a time value or range
\item one or more polarization states
\end{itemize}

The region on the sky should be something simple: a circle,
a range of coordinate values, or a polygon.

\item{Retrieve subsection of a 2D Image}

This is a special case of retrieving a subsection of a datacube,
where the cutout is only in the spatial axes.

\item{Retrieve subsection of a Spectrum}

This is a special case of retrieving a subsection of a datacube,
where the cutout is only in the spectral axis.

\item{Anticipated Future Use Cases. These use cases were taken into
account in the general design but remain to be developed and supported
in later versions of SODA.}

\begin{itemize}

\item{Provide the data in different formats}

Examples are images in PNG, or JPEG instead of FITS and spectra in csv,
FITS or VOTable.     

\item{Flatten a Datacube into a 2D Image}

\item{Reduce a Datacube into a 1D Spectrum}

\item{Rebin Data by a Fixed Factor}

\item{Reproject Data onto a Specified Grid}

\item{Compute Aggregate Functions over the Data}

\item{Apply Standard Function to Data Values}

It could be 
``denoising'' with standard methods or ``on the fly'' recalibration.

\item{Apply Arbitrary User-Specified Function to Data Values}

\item{Run Arbitrary User-Supplied Code on the Data}

\end{itemize}

\end{itemize}

\section{Resources}
\label{sec:resources}

SODA services are implemented as HTTP REST \citep{richardson07} web
services with a \{sync\} resource that conforms to the DALI-
sync resource description.

\begin{table}[ht]
\begin{tabular}{lll}
\sptablerule
\textbf{resource type}&\textbf{resource name}&\textbf{required}\cr
\sptablerule
\{sync\}&service specific&\tiny{(only one of \{sync\} or \{async\} required})\cr
\{async\}&service specific&\tiny{(only one of \{sync\} or \{async\} required})\cr
{DALI-examples}&/examples&no\cr
{VOSI-availability}&service specific&yes\cr
{VOSI-capabilities}&/capabilities&yes\cr
\sptablerule
\end{tabular}
\caption{Endpoints for SODA services}
\end{table}

A stand-alone SODA service may have one or both of the \{sync\} and
\{async\} resources. For either type, it could have multiple resources
(e.g. to support alternate authentication schemes). The SODA service may
also include other custom or supporting resources.

Either the \{sync\} or \{async\} SODA capability may be included as part
of other web services. For example, a single web service could contain
the SIA-2.0 \{query\} capability, the DataLink-1.0 \{links\} capability,
and the SODA \{sync\} capability. Such a service must also have the
VOSI-availability and VOSI-capabilities resources to report on and
describe all the implemented capabilities.

\subsection{\{sync\} resource}
\label{sec:sync}

The \{sync\} resource is a synchronous web service resource
that conforms to the DALI-sync description. Implementors
are free to name this resource however
they like, except that the name must consist of one URI segment only (i.e.,
contain no slash).  This is to allow clients, given the access URL, 
to reliably find out the URL of the capabilities endpoint. 
Clients, in turn, can find the resource path using the
VOSI-capabilities resource, but will in general be provided the access
URLs through a previous data discovery query or through direct user
input.

The \{sync\} resource performs the data access as specified by
the input parameters and returns the data directly in the
output stream. Synchronous data access is suitable when the
operations can be quickly performed and the data stream can
be setup and written to (by the service) in a short period
of time (e.g. before any timeouts).

\subsection{\{async\} resource}
\label{sec:async}

The \{async\} resource is an asynchronous web service resource
that conforms to the DALI-async description.  The considerations on
naming the resource given in sect.~\ref{sec:sync} apply here as well.

The \{async\} resource performs the data access as specified
by the input parameters and either (i) stores the results
for later transfer or (ii) pushes the results to a specified
destination (e.g. to a VOSpace location). Asynchronous data
access usually introduces resource constraints on the
service (which may be limited) and usually imposes a higher
latency before any results can be seen because the location
of results does not have to be valid until the data access
job is complete. Asynchronous data access is intended for
(but not limited to) use when the operations take
considerable time and results must be staged (e.g. some
multi-pass algorithms or operations that result in multiple
outputs).

\subsection{Examples: DALI-examples}
\label{sec:examples}

SODA services should provide a DALI-examples resource
with one example invocation that shows the variety of
operations the service can perform. Example operations using
the \{sync\} resource and that output a small data stream are
preferred, as the examples may be used by automatic validators doing
relatively frequent (of order daily) queries.

Parameters to be passed to the service must be given using the DALI
\texttt{generic-parameter} term.

\subsection{Availability: VOSI-availability}
\label{sec:availability}

A SODA web service must have a VOSI-availability
resource as described in DALI.

\subsection{Capabilities: VOSI-capabilities}
\label{sec:capabilities}

A web service that includes SODA capabilities must
have a VOSI-capabilities resource as described in DALI.
 The standardID for the \{sync\} resource is
$$\hbox{\texttt{ivo://ivoa.net/std/SODA\#sync-1.0}}$$

The standardID for the \{async\} resource is
$$\hbox{\texttt{ivo://ivoa.net/std/SODA\#async-1.0}}$$

All DAL services must implement the \texttt{/capabilities} resource.
The following capabilities document shows the minimal
metadata for a stand-alone SODA service and does not
require a registry extension schema:

\begin{lstlisting}[language=XML]
<?xml version="1.0"?>
<capabilities 
    xmlns:vosi="http://www.ivoa.net/xml/VOSICapabilities/v1.0" 
    xmlns:xsi="http://www.w3.org/2001/XMLSchema-instance" 
    xmlns:vod="http://www.ivoa.net/xml/VODataService/v1.1">
    
  <capability standardID="ivo://ivoa.net/std/VOSI#capabilities">
    <interface xsi:type="vod:ParamHTTP" version="1.0">
      <accessURL use="full">http://example.com/data/capabilities</accessURL>
    </interface>
  </capability>
  
  <capability standardID="ivo://ivoa.net/std/VOSI#availability">
    <interface xsi:type="vod:ParamHTTP" version="1.0">
      <accessURL use="full">
         http://example.com/data/availability
      </accessURL>
    </interface>
  </capability>
  
  <capability standardid="ivo://ivoa.net/std/SODA#sync-1.0">
    <interface xsi:type="vod:ParamHTTP" role="std" version="1.0">
      <accessURL use="full">
        http://example.com/data/sync
      </accessURL>
    </interface>
  </capability>
  
  <capability standardid="ivo://ivoa.net/std/SODA#async-1.0">
    <interface xsi:type="vod:ParamHTTP" role="std" version="1.0">
      <accessURL use="full">
        http://example.com/data/async
      </accessURL>
    </interface>
  </capability>
</capabilities>
\end{lstlisting}

Note that the \{sync\} and \{async\} resources do not have to be
named as shown in the accessURL(s) above. Multiple
interface elements within the \{sync\} and the \{async\} capabilities
may be included; this is typically used if they differ in
protocol (http vs. https) and/or authentication
requirements.

\section{Parameters for SODA \{sync\} and \{async\}}
\label{sec:parameters}

The \{sync\} and \{async\} resources accept the same set of
parameters.

\subsection{Parameter multiplicity}
Common methods of passing parameters to
HTTP services  allow passing the same parameter multiple times.
In the following, we call a parameter that is specified multiple times
in either way as having multiple values.

POL is a special case for multiplicity (see below)
 For all other parameters, support for multiple values of parameters is optional. 
If a request includes multiple values for a parameter and the 
service does not support multiple values for that parameter, the 
request must fail with the MultiValuedParamNotSupported error listed
below (\ref{sec:error-codes}). For example, a service may 
allow only single values for ID but multiple values for cutout parameters. 
Supported multiplicity may also differ between {sync} and {async} requests.

\enlargethispage{\baselineskip}

In general, services would support multi-valued parameters as they may be 
able to provide more efficient access to data files. Clients may attempt to use 
multi-valued parameters, but must be prepared to fall back to multiple requests 
if the service indicates this is not supported. A future version of
DataLink should provide a mechanism to describe parameter 
multiplicity.

\subsection{Common Parameters}

\subsubsection{ID}
\label{sec:ID}

The ID parameter is used to specify the dataset or file to
be accessed. The values for the ID parameter are generally
discovered from data discovery requests. The
values must be treated as opaque identifiers that are used
as-is. The DataLink specification describes mechanisms
for conveying opaque parameters and values in service
descriptor resources that can be used by clients to set the
ID parameter.

The UCD describing the ID parameter is
\ucd{meta.ref.url;meta.curation}.SODA

\subsection{Filtering Parameters}

Filtering parameters are used to extract subsets of large
datasets or data files. The extraction uses a best possible match to the requested subset. In case the parameter values excede the size of the archived dataset  the service operates a reduction of these values to the archived dataset size.

\subsubsection{POS}
\label{sec:POS}

The POS parameter defines the positional region(s) to be
extracted from the data. The value is made up of a shape
keyword followed by coordinate values. The
allowed shapes are given in Table~\ref{tab:shapetypes}.

\begin{table}[h]
\begin{tabular}{ll}
\sptablerule
\textbf{Shape}&\textbf{Coordinate values}\cr
\sptablerule
\texttt{CIRCLE}&\texttt{<longitude> <latitude> <radius>}\cr
\texttt{RANGE}&\texttt{<longitude1> <longitude2> <latitude1> <latitude2>}\cr
\texttt{POLYGON}&\texttt{<longitude1> <latitude1> ... (at least 3 pairs)}\cr
\sptablerule
\end{tabular}
\caption{POS Values in Spherical Coordinates}
\label{tab:shapetypes}
\end{table}

As in DALI intervals, open ranges use -Inf or +Inf as one limit.

\goodbreak
Examples for POS values:

\begin{itemize}
\item A circle at (12,34) with radius 0.5:

\begin{lstlisting}
POS=CIRCLE 12 34 0.5
\end{lstlisting}

\item A range of [12,14] in longitude and [34,36] in latitude:

\begin{lstlisting}
POS=RANGE 12 14 34 36
\end{lstlisting}

\item A polygon from (12,34) to (14,34) to (14,36) to (12,36) and
(implicitly) back to (12,34):

\begin{lstlisting}
POS=POLYGON 12 34 14 34 14 36 12 36
\end{lstlisting}

\item A band around the equator:

\begin{lstlisting}
POS=RANGE 0 360 -2 2
\end{lstlisting}

\item The north pole:

\begin{lstlisting}
POS=RANGE 0 360 89 90
\end{lstlisting}
\end{itemize}

All longitude and latitude values (plus the radius of the
CIRCLE) are expressed in degrees in ICRS.\footnote{A future
version of this specification may allow the use of other
reference systems (specifically the native system of the
data).}

The UCD  describing the POS parameter is \ucd{pos.outline;obs}.

Since it is string-valued, POS is unitless; however, the numeric values
contained in the string are all in decimal degrees. In VOTable, the
POS parameter has \verb|datatype="char"| and \verb|arraysize="*"|.

POS is included in SODA  for consistency with the SIA-2.0
 query parameter of the same name. Note that use of the
POS parameter with shape keyword ``CIRCLE'' provides the equivalent
spatial region as the CIRCLE parameter and POS with the shape keyword
``POLYGON'' is equivalent to the POLYGON parameter. There is no
type-specific parameter that is equivalent to the ``RANGE'' shape
keyword. There is no way for a service provider to declare support for a
subset of the POS shape keywords in a DataLink
service descriptor; either POS is included or not and if included then
all keywords must be supported.

\subsubsection{CIRCLE} 
\label{sec:CIRCLE}

The CIRCLE parameter defines a spatial region using the \xtype{circle}
xtype defined in DALI.

Example: a circle at (12,34) with radius 0.5:

\begin{lstlisting}
CIRCLE=12 34 0.5
\end{lstlisting}

The UCD describing the CIRCLE parameter is
\ucd{pos.outline;obs}.

CIRCLE is equivalent in functionality  to \texttt{POS=CIRCLE ...}.  Data type  and unit metadata are unambiguously defined.

\subsubsection{POLYGON}
\label{sec:POLYGON}

The POLYGON parameter defines a spatial region using the \xtype{polygon}
xtype defined in DALI.

Example: a  polygon from (12,34) to (14,34) to (14,36) to (12,36) and
(implicitly) back to (12,34):

\begin{lstlisting}
POLYGON=12 34 14 34 14 36 12 36
\end{lstlisting}

The UCD describing the POLYGON parameter is
\ucd{pos.outline;obs}.

POLYGON is equivalent in functionality to \texttt{POS=POLYGON ...}.  Data type  and unit metadata are unambiguously defined.

\subsubsection{BAND}
\label{sec:BAND}

The BAND parameter defines the wavelength interval(s) to be extracted from
the data using a floating point interval (\verb|xtype="interval"|) as
defined in DALI.  The value is an open or closed numeric
interval with numeric values interpreted as wavelength(s) in metres. As
in DALI, open intervals use -Inf or +Inf as one limit.

\begin{itemize}
\item The closed interval [500,550]:

\begin{lstlisting}
BAND=500 550
\end{lstlisting}

\item The open interval (-inf,300]:

\begin{lstlisting}
BAND=-Inf 300
\end{lstlisting}

\item The open interval [750,inf):

\begin{lstlisting}
BAND=750 +Inf
\end{lstlisting}

\item The scalar value 550, equivalent to [550,550]:

\begin{lstlisting}
BAND=550 550
\end{lstlisting}

\end{itemize}

Extracting using a scalar value should normally extract a
single pixel along the energy axis of the data; extracting
using an interval should extract one or more pixels.

All energy values are expressed as barycentric wavelength in
meters.\footnote{A future version of this specification may allow the
use of other reference systems (specifically the native
system of their data).}

The UCD describing the BAND parameter is \ucd{em.wl;stat.interval}.

\subsubsection{TIME}
\label{sec:TIME}

The BAND parameter defines the time interval(s) to be extracted from the
data using a floating point interval (\verb|xtype="interval"|) as
defined in DALI.  The value is an open or closed
interval with numeric values interpreted as Modified Julian Date(s) in
UTC. As in DALI, open intervals use -Inf or +Inf as one limit.

\begin{itemize}
\item An open interval from the MJD 55100.0 and all later times:

\begin{lstlisting}
TIME= 55100.0 +Inf
\end{lstlisting}

\item A range of MJD values:

\begin{lstlisting}
TIME=55123.456 55123.466
\end{lstlisting}

\item An instant in time using Modified Julian Date:

\begin{lstlisting}
TIME=55678.123456 55678.123456
\end{lstlisting}
\end{itemize}

The UCD describing the TIME parameter is
\ucd{time.interval;obs.exposure}.

\subsubsection{POL}
\label{sec:POL}

The POL parameter defines the polarization state(s) (Stokes)
to be extracted from the data.

\begin{itemize}
\item Extract the unpolarized intensity:
\begin{lstlisting}
POL=I
\end{lstlisting}
\item Extract the standard circular polarization:
\begin{lstlisting}
POL=V
\end{lstlisting}

\item Extract only the IQU components:
\begin{lstlisting}
POL=I
POL=Q
POL=U
\end{lstlisting}
\end{itemize}

As shown in the example above, the POL parameter must support multiple values 
for both \{sync\} and \{async\} requests.  Unlike general filtering parameters, 
all values of POL are combined into a single filter; for example, if the request
includes the three values above, the job would generate one result with
some or all of these polarization states (per combination of ID and
other filtering parameters).

The UCD  describing the POL parameter is
\ucd{meta.code;phys.polarization}.

\subsection{Filtering parameters and ObsCore data model}

Filtering parameters drive the generation of virtual datasets.  The ObsCore model is perfectly valid to  describe virtual data that SODA is able to generate. Hence all SODA filtering parameters are coupled with some Obscore model concepts.

The spatial parameters (CIRCLE, POLYGON and POS) constrain the spatial support of the output virtual dataset.

The TIME parameter constrains the time bounds of the SODA output virtual dataset. 

The BAND parameter constrains the spectral bounds of the SODA output virtual 
dataset. 

Support and bounds of the output datasets for each parameter are included in support and bounds of the archived dataset.

The POL parameter constrain the  list of polarization states in  the output virtual dataset. The
valid values for this param are included in the list given by the value
of the pol\_states attribute of the archived dataset.

\begin{table}[ht]
\begin{tabular}{l l l}
\sptablerule
\textbf{SODA}&\textbf{ObsCore utypes}&\textbf{ObsCore}\cr
\textbf{parameters}&&\textbf{attribute names}\cr
\sptablerule
\tiny{POS|CIRCLE|POLYGON}&obscore:Char.SpatialAxis.Coverage.Support.Area&s\_region\cr
BAND&obscore:Char.spectralAxis.Coverage.Bounds.Limits&\cr
TIME&obscore:Char.TimeAxis.Coverage.Bounds.Limits&\cr
POL&obscore:Char.PolarizationAxis.stateList&pol\_states\cr
\sptablerule
\end{tabular}
\caption{ObsCore utypes correspondance with standard SODA parameters}
\end{table} 

\subsection{Three-Factor Semantics}

Parameters in SODA are uniquely defined by the triple of name, UCD, and unit.  Data services are free to
support as many such parameters as is appropriate for their datasets, in
addition to supporting standard parameters.  With the three factors, it
is unlikely that two service providers will by accident use the same
three factors for parameters of differing semantics.  

To identify parameters, clients must use the three factors
name, UCD, and unit.  This is true for both the standard parameters
defined here and custom parameters introduced by services.  For
instance, a BAND parameter that is missing the em.wl;stat.interval UCD or has a
unit that is not meter must not be treated as the SODA BAND
parameter.
With standard parameters as defined in this document, clients can rely
on certain semantics and exploit that knowledge in the provision of
special UIs or APIs. 
  Standard parameters defined so far are given
in table~\ref{table:standardpars}.
For the time being, instructions for how to propose
additional  parameters will be given on the landing page of the IVOA
DAL working group\footnote{At the time of writing, this is\\
 \url{http://wiki.ivoa.net/twiki/bin/view/IVOA/DefiningServiceParameters}}

Table~\ref{table:freepars} is an exemple of definition of additional custom non-standard parameters.

\begin{table}[ht]
\begin{tabular}{l l l l}
\sptablerule
\textbf{Name}&\textbf{UCD}&\textbf{Unit}&\textbf{Semantics} \cr
\sptablerule
ID&meta.ref.url;meta.curation&&cf.~sect.~\ref{sec:ID} \cr
CIRCLE&pos.outline;obs&deg&cf.~sect.~\ref{sec:CIRCLE} \cr
POLYGON&pos.outline;obs&deg&cf.~sect.~\ref{sec:POLYGON} \cr
POS&pos.outline;obs&&cf.~sect.~\ref{sec:POS} \cr
BAND&em.wl;stat.interval&m&cf.~sect.~\ref{sec:BAND} \cr
TIME&time.interval;obs.exposure&d&cf.~sect.~\ref{sec:TIME} \cr
POL&meta.code;phys.polarization&&cf.~sect.~\ref{sec:POL} \cr
\sptablerule
\end{tabular}
\caption{Three-Factor Semantics for standard SODA parameters}
\label{table:standardpars}
\end{table}

\begin{table}[ht]
\begin{tabular}{l l l l}
\sptablerule
\textbf{Name}&\textbf{UCD}&\textbf{Unit}&\textbf{Semantics} \cr
\sptablerule
ID&meta.ref.url;meta.curation&& dataset identifier \cr
KERNSIZE&phys.size.radius&pixel& convolution kernel radius\cr
KERNTYPE&meta.code.class&&convolution kernel type \cr
&&& (Gaussian, Airy, etc...) \cr
\sptablerule
\end{tabular}
\caption{Example three-factor semantics for convolution-related custom parameters}
\label{table:freepars}
\end{table}

Both standard and non-standard parameters should follow DALI conventions
if at all possible.  Roughly, float-valued target fields should be accessed or
constrained via interval-valued parameters (i.e., do not split up
minimum and maximum into separate parameters).  Depending on their
semantics, integer parameters should either be intervals or enumerated
parameters (which typically can be repeated in the manner of POL). 
Geometry fields should be
accessed or constrained using geometry values (circle and polygon xtypes
from DALI), following the examples of CIRCLE
(\ref{sec:CIRCLE}) and POLYGON (\ref{sec:POLYGON}).

Parameter metadata, including three-factor semantics, is conveyed to
clients via DataLink service descriptor(s) as
described in Section~\ref{sec:integration}.

\section{Integration of Service Capabilities}
\label{sec:integration}

Finding and using SODA services depends on several other standards;
service providers can follow one or more strategies in integrating a
range of standard and custom services with their SODA implementation.
Here we describe these strategies and show how to use the standards
together.

Within the IVOA architecture, SODA services could be found and used in two
ways. First, a SODA service could be found in the IVOA Registry and used
directly. Second, a description of a SODA service may be found along
with specific dataset metadata; this is the primary anticipated usage:
clients discover applicable SODA services while doing data discovery
queries.

The DataLink recommendation provides a mechanism
to include ``a description of a SODA service'' using a standard resource
called a service descriptor. The service descriptor can be included in any
VOTable \citep{std:VOTable} output and can describe the parameters for
use with a DALI-sync or DALI-async compliant capability which may be a standard
 service or a custom service. Since the service descriptor can describe all input parameters,
it can declare available standard parameters, extensions (custom
parameters in standard services), and parameters for custom services.
This mechanism is expected to be the primary means for finding and using
a SODA service. 

A generic SODA sync service descriptor describing the standard
parameters (see sect.~\ref{sec:parameters}):

\begin{lstlisting}[language=XML]
<RESOURCE type="meta" ID="soda-sync" utype="adhoc:service">
  <PARAM name="standardID" datatype="char" arraysize="*" 
         value="ivo://ivoa.net/std/SODA#sync-1.0" >
   <DESCRIPTION>service protocol standard id</DESCRIPTION>
  </PARAM>
  <PARAM name="accessURL" datatype="char" arraysize="*" 
        value="http://example.com/soda/sync" >
  <DESCRIPTION>access url of the service</DESCRIPTION>
  </PARAM>
  <GROUP name="inputParams">
    <PARAM name="ID" ucd="meta.ref.url;meta.curation" 
            ref="idcolumn-ref" 
            datatype="char" arraysize="*" value="" >
    <DESCRIPTION>The publisher DID of the dataset of interest</DESCRIPTION>
    </PARAM>
    <PARAM name="POS" ucd="pos.outline;obs" 
            datatype="char" arraysize="*" value="" >
     <DESCRIPTION>Region to  cut out, as Circle, Box, or Polygon</DESCRIPTION>
    </PARAM>
    <PARAM name="CIRCLE" unit="deg" ucd="pos.outline;obs" 
            datatype="double" arraysize="3" 
            xtype="circle" value="" >
    <DESCRIPTION>A circle that should be covered by the cutout.</DESCRIPTION>
    </PARAM>
    <PARAM name="POLYGON"  unit="deg" ucd="pos.outline;obs"
            datatype="double" arraysize="*" 
            xtype="polygon"  value="" >
    <DESCRIPTION>A polygon  that should be covered by the cutout.</DESCRIPTION>
    </PARAM>
    <PARAM name="BAND" unit="m" ucd="em.wl;stat.interval" 
            datatype="double" arraysize="2" 
            xtype="interval" value="" >
    <DESCRIPTION>The wavelength intervals to be extracted</DESCRIPTION>
    </PARAM> 
    <PARAM name="TIME" ucd="time.interval;obs.exposure" unit="d" 
            datatype="double" arraysize="2" 
            xtype="interval" value="" >
     <DESCRIPTION>TIME Interval to be extracted in MJD</DESCRIPTION>
    </PARAM>
    <PARAM name="POL" ucd="meta.code;phys.polarization" 
            datatype="char" arraysize="*" value="" >
     <DESCRIPTION> Polarization states list to be extracted</DESCRIPTION>
    </PARAM>
  </GROUP>
</RESOURCE>
\end{lstlisting}

This service descriptor is generic because the ID parameter uses a
\xmlel{ref} attribute to specify that identifier values come from
elsewhere in the document (usually this refers to a FIELD element that
describes a table column within another RESOURCE element). Thus, this
descriptor can be used with any ID values in that column.

The PARAM with \verb|name="standardID"| specifies that this service is
a SODA sync service. The standardID values for SODA are specified in
Section~\ref{sec:capabilities}. 

The GROUP with \verb|name="inputParams"| shows the standard
description of the standard SODA parameters as defined in
Section~\ref{sec:parameters}. Services should only include parameter
descriptions for supported parameters; in a generic service descriptor
``supported'' means supported by the implementation and does not imply
that use of that parameter is applicable to all data (e.g. to all
possible identifier values).

All PARAMs in the descriptor may include a \xmlel{VALUES} subelement. This
element is  providing \xmlel{PARAMETER} domain limits or list of admitted
values. See section \ref{sec:disc-links-soda} for a full description of the usage of this feature.

\subsection{SODA Service Descriptor from Data Discovery}
\label{sec:disc-soda}

In the simplest case, the identifiers found via data discovery (e.g. the
\texttt{obs\_publisher\_did} in ObsCore) can be
used directly with an associated SODA service. Then the query response from
SIAv2  or TAP should include one or
more DataLink service descriptors that describe the
SODA capabilities. These would have a \texttt{standardID} parameter
specifying SODA \{async\} or SODA \{sync\} as specified in
Section~\ref{sec:capabilities} and an appropriate \texttt{accessURL}
parameter for the service. If the service is registered, the provider
can include a \texttt{resourceIdentifier} parameter which will contain 
the registered identifier of the service. 

The supported SODA
service parameters (standard and custom) would be declared in the
inputParams group of the service descriptor. 

The declaration of the ID parameter will specify which column in the
data discovery response contains the  suitable identifier; although this
is usually the obs\_publisher\_did from the ObsCore data model, this is
not required and the provider may have the ID parameter reference
another (possibly custom) column.

The data discovery response will in general contain metadata the client
can use to determine the values of SODA filtering parameters that will
yield valid subsets of the data. For example, standard data discovery
using either SIAv2 or TAP and ObsCore will provide metadata for
specifying POS, CIRCLE, and POLYGON (s\_region, s\_ra, s\_dec, s\_fov),
BAND (em\_min, em\_max), TIME (t\_min, t\_max), and POL (pol\_states)
parameters.

When a service descriptor for a SODA service is provided in the data
discovery response, it should be a generic descriptor (see above) for
use with multiple ID values. Thus, there will normally be a single
service descriptor for each available service.

\subsection{SODA Service Descriptor from DataLink}
\label{sec:disc-links-soda}

The alternative scenario has the discovery service return Datalink
documents (see DataLink for the two ways to do that:  via the
access\_url or via a DataLink ``service descriptor'' in 
the query response).  These
Datalink documents can then contain one or more SODA descriptor(s),
most typically one per dataset described.  To allow SODA clients
the inference of parameter ranges and the presentation of useful
user interfaces, data providers SHOULD communicate the admissable
ranges of the parameters in question using the VOTable
\xmlel{VALUES} element.

For float-valued intervals (e.g., the standard BAND and TIME
parameters), \xmlel{VALUES/MIN} and \xmlel{VALUES/MAX} should be used to
communicate the range of values for which clients can expect to
receive data.  Example:

\begin{lstlisting}[language=XML]
  <PARAM name="BAND" unit="m" ucd="em.wl;stat.interval"
    datatype="double" arraysize="2"      
    xtype="interval" value="">     
    <DESCRIPTION>The wavelength intervals to be extracted</DESCRIPTION>
    <VALUES>                                                           
      <MIN value="3e-7"/>
      <MAX value="8e-7"/>
    </VALUE>             
  </PARAM>  
\end{lstlisting}

Enumerated values, both for integral and textual types, use
\xmlel{VALUES}/\xmlel{OP\-TION} elements unless there are too many possible
values.  Again, only values for which nonempty responses can be
expected for the described dataset should be listed.  Example:

\begin{lstlisting}[language=XML]
  <PARAM name="POL" ucd="meta.code;phys.polarization"
    datatype="char" arraysize="*" value="">          
    <DESCRIPTION>Polarization states to be extracted.</DESCRIPTION>
    <VALUES>
      <OPTION>I</OPTION>
      <OPTION>V</OPTION>
    </VALUE>
  </PARAM>
\end{lstlisting}

In case the option enumeration becomes too large, the description
of the parameter should carefully describe what values are
admissable, e.g., by providing a link to an enumeration in the
\xmlel{DESCRIPTION}.

Intervals of integers are described analogous to float-valued
intervals, i.e., using \xmlel{MIN} and \xmlel{MAX} elements.

Standard VOTable semantics are insufficient for the metadata of the SODA
POLYGON and CIRCLE parameters.  We therefore define special cases for
the \xmlel{xtype}s \emph{circle} and \emph{polygon} at least until such
time when a proper data model for space-time coordinates will define a
different way to communicate such coverages within VOTables.

For CIRCLE, only a \xmlel{MAX} is given. It contains three
floating point values, separated by whitespace.  These correspond
to the RA and Dec of the center of a spherical circle covering the
dataset, and a radius of such a covering circle.  Data providers
SHOULD make sure they choose the center and radius such that the
covering circle is close to the minimal one of the dataset.
Example:

\begin{lstlisting}[language=XML]
  <PARAM name="CIRCLE" unit="deg" ucd="pos.outline;obs"
    datatype="double" arraysize="3"
    xtype="circle" value="">
    <DESCRIPTION>
      A spherical circle to be contained by the cutout
    </DESCRIPTION>
    <VALUES> <MAX value="12.0 34.0 0.5"/> </VALUES>
  </PARAM>
\end{lstlisting}

For POLYGON, again only a \xmlel{MAX} is given.  It consists of
a sequence of floating-point values, again separated by blanks,
describing RA and Dec of the vertices of a spherical polygon
covering the dataset.  Data providers are encouraged to choose a
minimal polygon.  Example:

\begin{lstlisting}[language=XML]
<PARAM name="POLYGON"  unit="deg" ucd="pos.outline;obs"
        datatype="double" arraysize="*"
        xtype="polygon"  value="">
  <DESCRIPTION>A polygon to be contained by the cutout</DESCRIPTION>
  <VALUES>
    <MAX value="11.5 33.5 12.5 33.5 12.5 34.5 11.5 34.5"/>
  </VALUES>
</PARAM>
\end{lstlisting}

Angles in both CIRCLE and POLYGON are in degrees.  As in the input,
the ICRS reference system is assumed, with no further metadata (e.g.,
reference position) prescribed by this standard.  Further metadata 
should be given using standard STC annotation when the formalism to do
that is finalised.

For POS, useful metadata cannot be given.  Services supporting POS
should therefore provide POLYGON as well, and clients wishing to
use POS should infer sensible values for that parameter from
\xmlel{VALUES} given for POLYGON.

A full example for a dataset-specific datalink descriptor is given in
appendix~\ref{app:fullsoda}.
 
Providing values in the parameter descriptions of a data-specific
service descriptor implies that the resource generating this has access
to the applicable metadata. Depending on system architecture, this may be
difficult to implement.\footnote{An ``autodescription'' mechanism where the SODA
service can generate a data-specific service descriptor of itself
may be included in SODA-1.1 or later.}

\subsection{Finding a SODA Service in the Registry}
\label{sec:reg-soda}

Resources in the IVOA Registry may include SODA capabilities. However, 
in order to
use such services, clients require prior knowledge of suitable
identifiers that are usable with a registered SODA service. As a result,
finding and
using a SODA service via the registry is not expected to be a common
usage pattern.

\section{\{sync\} Responses}

All responses from the \{sync\} resource follow the rules for
DALI-sync resources, except that the \{sync\} response allows
for error messages for individual input identifier values.

\subsection{Successful Requests}

Successfully executed requests should result in a response
with HTTP status code 200 (OK) and a response in the format
requested by the client or in the default format for the
service.

If the values specified for cutout parameters do not include
any pixels from the target dataset/file, the service must
respond with HTTP status code 204 (No Content) and no
response body, as stated in DALI.

The service should set the following HTTP headers to the
correct values where possible.

\begin{tabular}{ll}
\sptablerule
Content-Type&media type of the response\cr
Content-Encoding&encoding/compression of the response (if applicable)\cr
\sptablerule
\end{tabular}

Since the response is usually dynamically generated, the
Content-Length and Last-Modified headers cannot usually be
set.

\subsection{Errors}
\label{sec:error-codes}

The error handling specified for DALI-sync resources applies
to service failure. Error codes are specified in DALI. Error documents should be text using the
text/plain content-type and the text must begin with one of
the following strings:

\begin{table}[h]
\begin{tabular}{l l}
\sptablerule
\textbf{Error Code} & \textbf{Description}  \cr
\sptablerule
Error&General error (not covered below) \cr
AuthenticationError&Not authenticated \cr
AuthorizationError&Not authorized to access the resource \cr
ServiceUnavailable&Transient error (could succeed with retry) \cr
UsageError&Permanent error (retry pointless) \cr
MultiValuedParamNotSupported&request included multiple values for a parameter\cr
&but the service only supports a single value \cr
\sptablerule
\end{tabular}
\caption{error messages with their meaning}
\end{table}

\section{\{async\} Responses}

The \{async\} resource conforms to the DALI-async resource
description, which means the job is a UWS job with all the
job control features available. All result files are to be
listed as children of the UWS results resource. The service
provider is free to name each result.

When multiple values of input parameters are accepted, 
each combination of values produces one result. For
example, if an \{async\} job included two CIRCLE and two BAND
values, there must be four results. If a combination
of input parameters does not produce a result (e.g. there is no 
overlap between the parameter values and data extent), the job results 
must contain a result entry that indicates this. This should be
a result URL which returns a text/plain document with a message
starting with one of the error labels in Section~\ref{sec:error-codes} 
above.

\appendix

\section{Full SODA Descriptor example}
\label{app:fullsoda}

Below is an example illustrating how a SODA descriptor for a dataset as
delivered in a DataLink document might look like (see
sect~\ref{sec:disc-links-soda}).  Note in particular how \xmlel{value}
is used in the declaration of the ID parameter to convey the fixed value
corresponding to the dataset described.

The particular dataset described here is a spectral cube.  Therefore
no TIME and POL parameters are defined.

The example also illustrates how a custom parameter (here, KIND) would
be declared.

\begin{lstlisting}[language=XML,basicstyle=\footnotesize]
  <RESOURCE ID="referenced" type="meta" utype="adhoc:service">
    <GROUP name="inputParams">
      <PARAM arraysize="*" datatype="char" name="ID" ucd="meta.id;meta.main" 
        value="ivo://org.gavo.dc/~?califa/datadr3/COMB/NGC0180.COMB.rscube.fits">
        <DESCRIPTION>The publisher DID of the dataset of interest</DESCRIPTION>
      </PARAM>
      <PARAM arraysize="*" datatype="char" name="POS" ucd="pos.outline;obs" 
        value="">
        <DESCRIPTION>Region to (approximately) cut out, as Circle, Box, 
          or Polygon</DESCRIPTION>
      </PARAM>
      <PARAM arraysize="*" datatype="double" name="POLYGON" 
        ucd="pos.outline;obs" unit="deg" value="">
        <DESCRIPTION>A polygon (as a flattened array of ra, dec pairs) that 
          should be covered by the cutout.</DESCRIPTION>
        <VALUES>
          <MAX value="9.499 8.626 9.499 8.645 9.478 8.645 9.478 8.626"/>
        </VALUES>
      </PARAM>
      <PARAM arraysize="3" datatype="double" name="CIRCLE" 
        ucd="pos.outline;obs" unit="deg" value="">
        <DESCRIPTION>A circle (as a flattened array of ra, dec, radius) 
          that should be covered by the cutout.</DESCRIPTION>
        <VALUES>
          <MAX value="9.4889955890 8.6358711588 0.0146493214"/>
        </VALUES>
      </PARAM>
      <PARAM arraysize="2" datatype="double" name="BAND" ucd="em.wl;stat.interval" 
        unit="m" value="" xtype="interval">
        <DESCRIPTION>Vacuum wavelength limits</DESCRIPTION>
        <VALUES>
          <MIN value="3.701e-07"/>
          <MAX value="7.501e-07"/>
        </VALUES>
      </PARAM>
      <PARAM arraysize="*" datatype="char" name="KIND" ucd="" value="">
        <DESCRIPTION>Set to HEADER to retrieve just the primary header, 
          leave empty for data.</DESCRIPTION>
        <VALUES>
          <OPTION name="Retrieve header only" value="HEADER"/>
          <OPTION name="Retrieve the full data, including header (default)" 
            value="DATA"/>
        </VALUES>
      </PARAM>
    </GROUP>
    <PARAM arraysize="*" datatype="char" name="accessURL" ucd="meta.ref.url" 
      value="http://dc.g-vo.org/califa/q3/dl/dlget"/>
    <PARAM arraysize="*" datatype="char" name="standardID" 
      value="ivo://ivoa.net/std/SODA#sync-1.0"/>
  </RESOURCE>
\end{lstlisting}

\section{Changes from Previous Versions}

\subsection{Changes from PR-SODA-20160429}
\begin{itemize}
\item Make multiple values for all parameters optional in both {sync} and 
{async} requests and introduce a specific error message if multiplicity of 
a parameter is not supported.
\item Added section introducing the different usage scenarios for SODA
and how they can interact with other DAL capabilities. Moved the bulk of
the normative text to an integration section so that it follows the
primary specification of SODA resources and parameters.
\item Re-organised so that UCDs for parameters are only specified once
in the section on three-factor semantics.
\item Added CIRCLE AND POLYGON ``double array'' parameters. POS is
retained for consistency with SIA-2.0 query. 
\item Interval xtype as strict arraysize=2 array consistently with DALI 1.1 
\item SODA autodescription is postponed to version 1.1.
\item VALUES for xtype=interval now use MIN and MAX rather than MAX
alone.
\end{itemize}

\subsection{Changes from WD-SODA-1.0-20151212}
\begin{itemize}
\item POS is now unitless
\item Aligned parameter UCDs with what is in ObsCore
\item Removed gratuitous xtypes.
\end{itemize}

\subsection{Changes from WD-SODA-1.0-20151120}

Change the name of the protocol. Suppression of SELECT and COORD. xtype description are in DALI. Reference to this has been added.

\subsection{Changes from WD-AccessData-1.0-20151021}

Added general introduction on PARAMETER description to
section 3. Modified SELECT and COORD sections in order to
detach them from SimDal. Added Appendix on xtype description
with BNF syntax.

\subsection{Changes from WD-AccessData-1.0-20140730}

\begin{itemize}
\item Removed REQUEST parameter since the DAL-WG decision to not
include it when there is only one value.

\item Clarified that ID and filtering parameters are single
valued for \{sync\} and multi-valued for \{async\}, with POL
being multi-valued but still being treated as a single
filter.
\end{itemize}

\subsection{Changes from WD-AccessData-1.0-20140312}

This is the initial document version.

\bibliography{ivoatex/ivoabib}

\end{document}